# Explainable artificial intelligence (XAI) for scaling: An application for deducing hydrologic connectivity at watershed scale

Sheng Ye[1], Jiyu Li[1], Yifan Chai[1], Lin Liu[2], Murugesu Sivapalan[4,5], Qihua Ran[2,3]*

[1]Institute of Water Science and Engineering, College of Civil Engineering and Architecture, Zhejiang University, Hangzhou 310058, China
[2]Key Laboratory of Hydrologic-Cycle and Hydrodynamic-System of Ministry of Water Resources, Hohai University, Nanjing 210098, China
[3]National Cooperative Innovation Center for Water Safety & Hydro-Science, Joint International Research Laboratory of Global Change and Water Cycle, Hohai University, Nanjing 210098, China
[4]Department of Civil and Environmental Engineering, University of Illinois at Urbana‐Champaign, Urbana, IL, USA
[5]Department of Geography and Geographic Information Science, University of Illinois at Urbana‐Champaign, Urbana, IL, USA

*Correspondence to*: Qihua Ran (ranqihua@hhu.edu.cn)

**Abstract.** Explainable artificial intelligence (XAI) methods have been applied to interpret deep learning model results. However, applications that integrate XAI with established hydrologic knowledge for process understanding remain limited. Here we show that XAI method, applied at point-scale, could be used for cross-scale aggregation of hydrologic responses, a fundamental question in scaling problems, using hydrologic connectivity as a demonstration. Soil moisture and its movement generated by physically based hydrologic model were used to train a long short-term memory (LSTM) network, whose impacts of inputs were evaluated by XAI methods. Our results suggest that XAI-based classification can effectively identify the differences in the functional roles of various sub-regions at watershed scale. The aggregated XAI results could be considered as an explicit and quantitative indicator of hydrologic connectivity development, offering insights to hydrological organization. This framework could be used to facilitate aggregation of other geophysical responses to advance process understandings.

## 1 Introduction

Deep learning (DL) refers to a class of machine learning algorithms in which a hierarchy of neural network layers are used to transform input data into more abstract and composite representations or patterns. It has been applied in hydrology for predictions over the past decade (Jiang et al 2022; Kratzert et al., 2018; Li et al., 2021; Shen et al., 2023; Sun et al., 2019; Wang et al., 2023). Along with the impressive achievements of DL, there is also increasing demand for transparency: what is the physical explanation of the DL model structure and can DL help advance scientific understanding (Blöschl et al., 2019; Preece et al., 2018; Shen et al., 2018). Explainable artificial intelligence (XAI) is a set of processes and methods that help human users to comprehend and gain trust in the output patterns created by machine learning (including DL) algorithms. In this way, it has been suggested that XAI can provide transparency and interpretability to DL models (Samek et al., 2021). Recently, XAI methods have been applied in hydrology to gain scientific insights (Jung et al., 2024; Pradhan et al., 2023; Topp et al., 2023; Wu et al., 2023). For example, Jiang et al. (2022) analyzed flood data across the United States to identify the dominant flooding mechanisms; Pradhan et al. (2023) used an XAI model for spatial flood susceptibility mapping.

However, most of the XAI applications to date have focused on the quantification of the overall importance of the influencing factors, and it has been rather difficult to decode a physical process from the pure mathematical model outputs (Nearing et al., 2021). Instead of being applied to the system like a lumped model, when applied at the grid scale like a distributed model using neighboring grid points as input features, XAI method could be used to evaluate the strength of links between neighboring grid points. In other words, the XAI results could provide a quantification of functional dependence of the dynamic responses at the grid scale, which can then be considered as a combination of a local or point value and the underlying organizational pattern that connects to neighboring grid points. This is the essential requirement for derivation of hydrologic responses at watershed or macro-scale from process descriptions at the local grid or micro-scale: a framework that builds on knowledge and interpretation of a pattern of spatial organization (Blöschl and Sivapalan 1995; Viney and Sivapalan 2004). Thus, when applied at the grid-scale, XAI method can evaluate the functional roles of grid-scale dynamics, which could be aggregated across the watershed to derive an indicator of watershed-scale response, providing an alternative approach to the upscaling of hydrologic responses. This is the motivation for this paper.

Hydrologic connectivity is chosen here as a demonstration of this aggregation framework. It is a watershed scale, oftentimes hidden, dynamic internal organizational feature, which characterizes the degree of connectivity between subregions within the watershed, including how it might change over time. It plays a central role in governing the dynamics of transport of water, solutes, organisms and energy, and how they might change with increasing spatial scale (Blume and van Meerveld, 2015; Saffarpour et al., 2016), contributing to the nonlinear (e.g., threshold like) dynamics in observed streamflow responses and contaminant breakthrough curves (van Meerveld et al., 2015; Wainwright and Bracken, 2011). However, hydrologic connectivity is difficult to measure directly and can only be inferred from observations of dynamic patterns of other variables such as soil moisture and/or groundwater table (Kiewiet et al., 2020; Pavlin et al. 2021; Zhang et al., 2021). There is clearly a difference in scales between distributed measurements (at the local or grid scale) of these inference variables and hydrologic connectivity, which is defined and meaningful only at a larger scale (hillslope or watershed scale). Field studies have indicated inconsistency in the dynamics of hydrograph and soil moisture (Tromp-van Meerveld and McDonnell, 2006). While a distributed physically based hydrologic model based on the Richards equation (Richards 1931) simulates soil water movement at each grid point within a watershed, the organized pattern of the water movement at the watershed scale, e.g., hydrologic connectivity, is not enabled by a priori model design but emerges in a self-organizing manner (Bracken et al., 2013). How variability at the local or micro-scale would manifest in or give rise to hydrologic connectivity, an emergent pattern or dynamics at the watershed or macro-scale is a question that is fundamental to modeling of flow and transport at the watershed scale and requires deeper investigation (Bracken et al., 2013).

These characteristics of hydrologic connectivity make it suitable for the test of our framework. In this study, we use the XAI method to aggregate point-scale interactions to watershed scale hydrologic response. Based on the hydrologic understanding gained, we show that the group aggregated XAI results have corresponding physical meanings and can be used to indicate the development of hydrologic connectivity. The goal of the study is to demonstrate that XAI, integrating with hydrologic knowledge, can be used to bridge across scales to advance scientific understanding.

## 2 Methodology

We first apply a distributed hydrologic model to the study watershed to generate event scale data on soil moisture distribution and its dynamics (i.e., soil moisture movement in three dimensions), which we then use to train a LSTM network to predict soil water movement at each grid. The XAI method is then applied to calculate the impacts of input features in the prediction. The importance of surrounding grids is classified and aggregated to watershed scale. Integrated with hydrologic knowledge, we show that the aggregated results could be interpreted as an indicator of the development of hydrologic connectivity (Fig. 1).

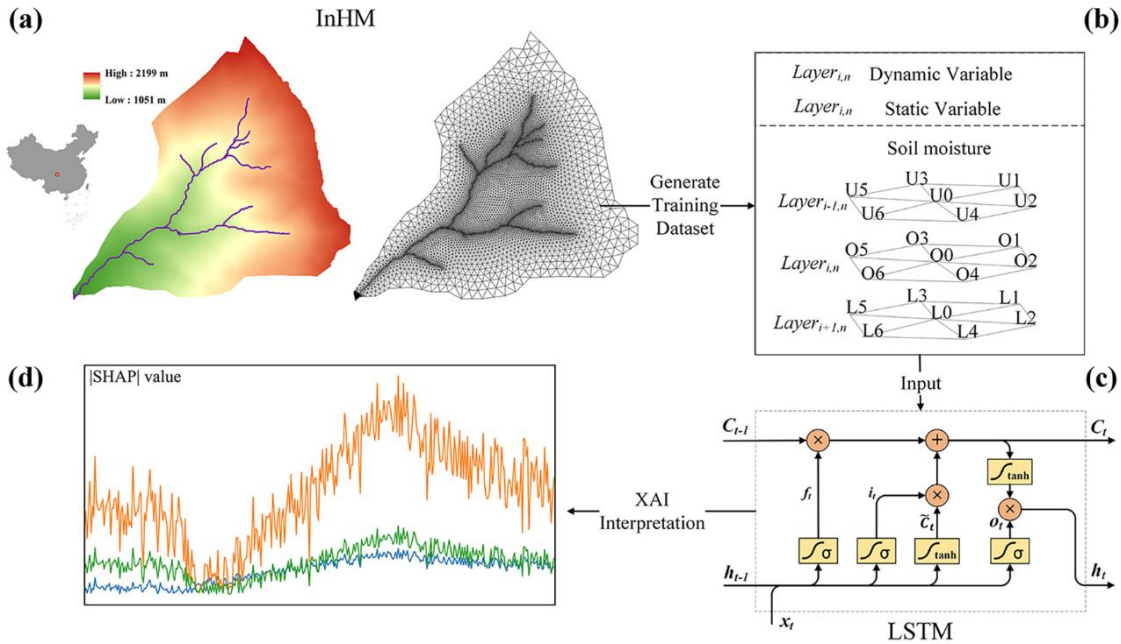

**Figure 1: Diagram of framework of this study. (a) Geographical location, elevation map, and the two-dimensional mesh grid of the study watershed for the physical model (InHM); (b) Input variables of LSTM, the six adjacent nodes (U1-U6, L1-L6, O1-O6) at each layer were ordered by elevation from high to low; (c) the model structure of LSTM; (d) interpretation results from XAI method.**

### 2.1 Study area & data

The study watershed, Jianpinggou (JPG), is a headwater watershed within the Yangtze River basin (Fig. 1a). It covers an area of 3.5 km2, with elevations ranging from 1051 meters to 2199 meters and topographic gradient varying from 0° to 63.3°, and averaging 32.4° (Ran et al., 2015). The mean annual precipitation is 1134mm, while the mean annual evaporation is around 900mm (Ye et al., 2023). 5-m digital elevation model (DEM) was digitized from topographic maps. The event rainfall data was recorded every minute using an automatic rainfall gauge and the event runoff data was measured at the outlet using a large-scale particle image velocimetry (LSPIV) system every 45 min in 2018 and 2019 (Ye et al., 2023).

**2.2 Distributed hydrologic model setup**

A physically based distributed model, Integrated Hydrology Model (InHM), which characterizes variably saturated flow in the subsurface continua based on the three-dimensional Richards' equation, was selected for this study (Vanderkwaak, 1999). It was chosen for the capability to simulate soil moisture content and 3-D soil water movement with high spatial-temporal resolution and full physical basis. The soil depth was set to 13 meters, and was divided into 19 layers in total, where the top 10 layers falling within the top one meter of soil were used to focus on root zone soil moisture dynamics. The porosity and saturated hydraulic conductivity ($K_s$) were determined from previous field measurements (Ye et al., 2023). The model was calibrated and validated at the event scale in previous work (Liu et al., 2021). A 12-day flow event was simulated to generate data capturing the event scale development of hydrologic connectivity and to train the deep learning network model. Details about InHM is shown in Text S1 (Ran et al., 2007).

**2.3 Deep learning method**

This study employed a single-layered chain-structured LSTM network to predict soil water movement in three dimensions. The selected input features are soil moisture of the prediction node (O0) and the surrounding nodes, precipitation, and static soil properties which only vary with depth (i.e., $K_s$, depth and soil layer thickness) (Fig. 1b). Data from 3203 nodes during 17,280 time steps at layers within top one meter were used for DL model simulation. To avoid confusion between surface runoff and subsurface soil movement in Layer 1 (0 cm), our predictions were conducted from Layer 2 (10 cm) to Layer 10 (90 cm). All features were normalized before training, and the temporal resolution was a minute. The output data was soil moisture movement velocities in three directions, Vx, Vy, and Vz (Fig. 1b), which were modeled separately. Nine days of the flow events were used for training, and the remaining three days were used for testing. The model was optimized by the Adam optimizer, with hyperparameter tuning conducted through the Asynchronous Successive Halving Algorithm (ASHA) with early stopping to enhance training efficiency. Mean Square Error (MSE) was used for the loss function and the Nash-Sutcliffe efficiency (NSE) coefficient was calculated for performance evaluation.

**2.4 The explainable AI method**

Expected Gradients, which combines Integrated Gradients, SmoothGrad and the SHapley Additive exPlanations (SHAP), was adopted for interpretation (Lundberg & Lee, 2017). It calculates the average importance of each input feature at each time step, i.e., how much each feature contributes to a model's prediction, by averaging across many possible scenarios (Erion et al, 2021). In this study, it was calculated with the SHapley Additive exPlanations (SHAP) package (Lundberg &

Lee, 2017) to quantify the impacts of each input feature on model prediction. Details about the calculation of Expected Gradients is shown in Text S2. The results will be referred to as SHAP values to represent the importance of input features. By quantifying the importance of neighboring grids in the prediction, SHAP values take into account of the functional roles of the dynamic responses at the grid scale. Therefore, the SHAP values could be considered as a combination of the point value and the underlying organizational pattern, making it suitable for cross-scale aggregation.

**2.5 Statistical method for classification**

K-means clustering algorithm was employed to classify the nodes based on temporal sequences of SHAP values. Dynamic Time Warping (DTW) was utilized as the distance metric to measure similarity between time series (Sakoe & Chiba, 1978). The DTW distance from sequences $[s_1^{(1)}, s_1^{(2)}, ... s_1^{(n)}]$ to $[s_2^{(1)}, s_2^{(2)}, ... s_2^{(m)}]$ can be expressed by the following Eq. (1):

$$DTW = \min_{\pi} \sqrt{\sum_{(i,j)\epsilon\pi}(s_1^{(i)} - s_1^{(j)})^2} \quad (1)$$

where $\pi = [\pi_0, \pi_1, ..., \pi_K]$ is a list of index pairs $\pi_K = (i_k, j_k)$ with $0 \leq i_k \leq n$ and $0 \leq j_m \leq m$. The classification was implemented using DTW Barycenter Averaging (DBA) from the *tslearn* package (Petitjean et al., 2011; Tavenard et al., 2020). A three-cluster solution was selected based on its relatively high silhouette scores and better representation of the physical heterogeneity within the watershed (Rousseeuw, 1987). The cluster centroid which represents the average sequence of cluster is used to derive the aggregated SHAP values at higher spatial scale, i.e., watershed scale.

**3 Results**

The NSE values for the prediction of soil water movement in three directions (Vx, Vy, Vz) were found to be reasonable, with most of them being higher than 0.8 (Fig. S1). Especially for the prediction of Vz, the NSE values for Layer 2 and Layer 3 were as high as 0.9 with only minor variations. This suggests that the LSTM model is able to capture meaningful patterns from the input data, providing reliable predictions for further interpretation.

The feature importance, quantified as SHAP values, was calculated for one flow event during the testing period as an example. All nodes were classified into three categories based on the sequences of SHAP values. Take Layer 2 as example, the averaged SHAP values of the three classes show distinct characteristics (Fig. 2). One class (orange dots) exhibits lower soil moisture level and lateral movement (Vx, Vy), along with higher vertical infiltration (Vz). The blue dot class maintains high soil moisture levels with little variation in lateral movement, while green dot class displays comparably high soil moisture level but exhibits greater increment in lateral movements. Given the temporal variation of the soil moisture and water movement, one may hypothesize that these three classes correspond to three sub-regions during runoff generation: the unsaturated, saturated, and riparian area. The spatial distribution of these three classes aligns with the regions identified by the numerical model, confirming our interpretation: the blue dots are in the saturated area along the river channel, the green ones locate mostly in the riparian area which reached saturation after rainfall, and the orange ones scatter along the unsaturated hillslopes (Fig. 3). Thus, we would name the three classes as: Channel Proximity nodes (CH, blue dots), Riparian Area nodes (TR, green dots), and Hillslope nodes (HL, orange dots).

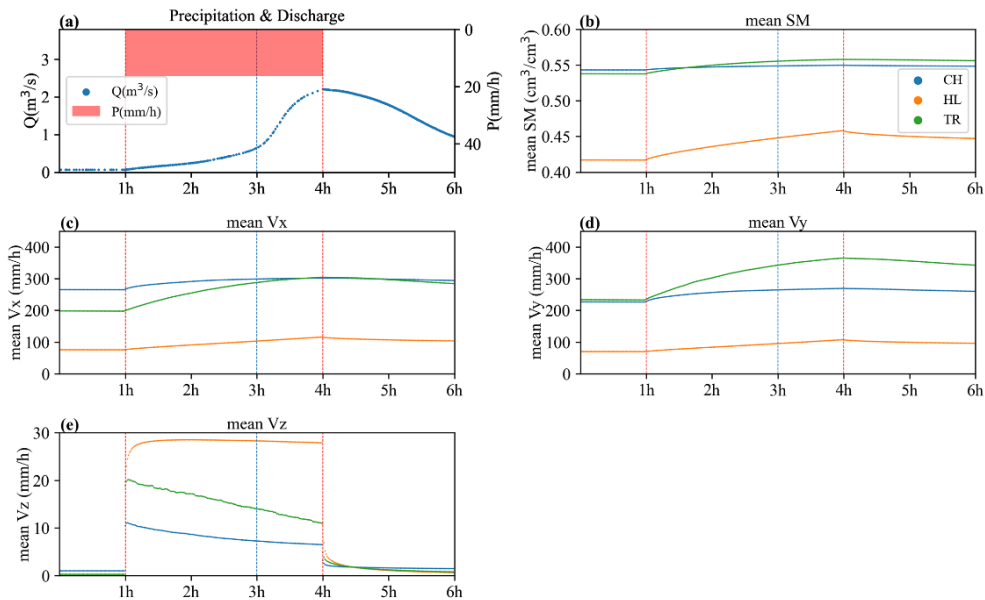

Figure 2: (a) The input precipitation and InHM simulated runoff results of the rainfall event used for interpretation, the red dash line represents the beginning and end of rainfall, and the blue dash line indicates the abrupt rise in runoff. (b) The class averaged soil moisture; and soil moisture dynamics at three directions (c) Vx, (d) Vy, and (e) Vz of Layer 2.

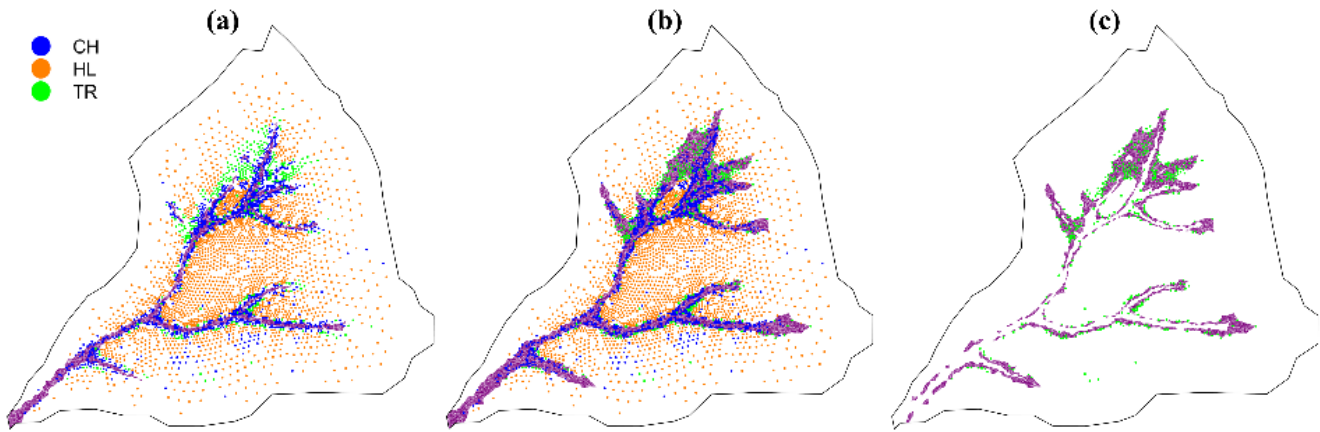

Figure 3: Saturated area at the (a) start of rainfall; (b) end of rainfall, and (c) the expanded area of saturation; the blue dots represent CH nodes, green dots are TR nodes, and orange ones are HL nodes.

The average sequences of SHAP values from cluster centroid were derived for input features to represent their aggregated impacts on vertical (Vz) and lateral soil water movement (Vx, Vy) at watershed scale. For the lateral movement simulation,

Vx, Vy (Fig. 4, S2), the average SHAP values of TR nodes showed distinct patterns from the CH and HL nodes, especially for the SHAP values of soil moisture. The influence of soil moisture was small and changed gradually at CH and HL nodes, while a drastic increase was found at TR nodes. The importance of soil moisture remained low at the early stage of rainfall events, until about two hours after the rainfall started, the impact of soil moisture began a rapid climb. This was around the time when the streamflow started an abrupt rise at the outlet (Fig. 2a). The timing of the jump varied with different soil depths, and was much less significant in the vertical simulation (Fig. S5 – S7). For the vertical simulation (Vz), the impact of rainfall was overwhelming at the HL nodes, which is consistent with our understanding of the importance of rainfall at hillslope scale.

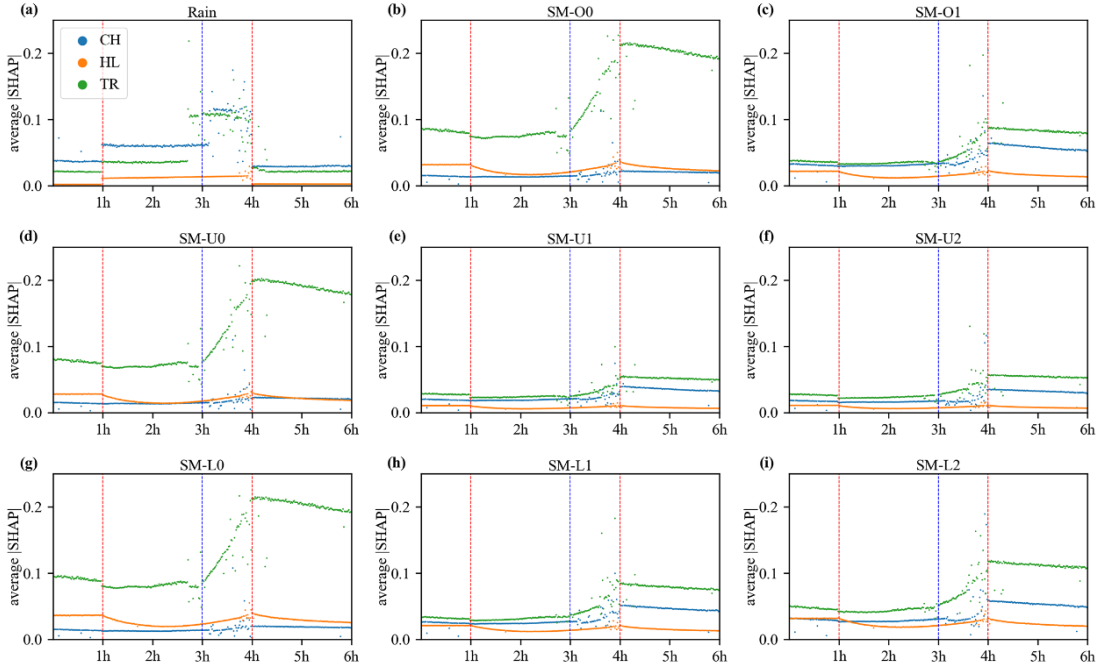

Figure 4: The average sequences of SHAP values of each class for rainfall and soil moisture at nodes with higher importance (O0, O1, U0, U1, U2, L0, L1, L2) in X-direction of Layer 2.

## 4 The potential of XAI in advancing scientific understanding

### 4.1 The development of hydrologic connectivity at watershed scale

Hydrologic connectivity plays an important role in runoff generation (Sivapalan, 2003). However, it can only be inferred from soil moisture and groundwater level indirectly, which need to be aggregated to derive watershed scale patterns that are comparable with hydrograph (Jencso et al., 2010; McGlynn and McDonnell 2003; Pavlin et al 2021; Zimmer and McGlynn 2017). Here, we used the SHAP values to aggregate dynamics of soil water movement from point scale to watershed scale. By quantifying the importance of surrounding grid points, the SHAP values capture the functional dependence at grid scale, thus can be considered as a combination of both point value and the underlying organizational pattern. Furthermore, when averaged within cluster, one may hypothesize that the aggregation of SHAP values could be used to indicate hydrologic connectivity at watershed scale. We examined this from two perspectives: (1) does classification based on SHAP values divide nodes into groups with physical meanings at watershed scale, and (2) could the average SHAP values generate results comparable at watershed scale to indicate hydrologic connectivity?

The spatial distribution of the three classes supported the hypothesis that that groups classified by SHAP values were consistent with the sub-regions derived at watershed scale from InHM for runoff generation: the unsaturated, riparian, and saturated area (Fig. 3). The synchronicity of the abrupt rise in the SHAP values of surrounding grids and the surge in streamflow suggested that the aggregated SHAP values could derive results that are comparable with hydrograph at watershed scale (Fig. 4). The sudden elevated soil moisture importance in TR nodes for the lateral movement prediction could be considered as the sign of activation of abundant soil water movement across riparian area, indicating the establishment of hydrologic connectivity at the watershed scale. The spatial distribution of surface water depth before and after the abrupt rise extracted from the hydrologic model simulation also helped validate our hypothesis: the streamflow in the main channel was not fully connected until the occurrence of the abrupt rise in SHAP values (Fig. S4).

Threshold behavior was found in the development of hydrologic connectivity: only when this threshold was exceeded, did the lateral movement of soil moisture became significantly active. Before the threshold was reached, the various parts of the whole watershed were not fully connected, while there was lateral movement at each node, yet that movement was concentrated at the local scale, and was not continuous at watershed scale. Once the threshold was reached, hydrologic connectivity was fully established across the nodes, the local scale movement at various places linked up together to form continuous flow pathways within the watershed from upstream to the watershed outlet, resulting in the rapid rise in hydrograph. Such threshold behavior has also been identified in previous observations at our study watershed: the shift from slow to fast flow occurred when the antecedent soil moisture and rainfall reached certain thresholds (Ye et al 2023; Zhang et al 2021), which further confirmed our hypothesis.

This pattern of abrupt rise was most apparent in the riparian areas, which have been identified as the key areas in the watershed to understand the hydrologic connection between hillslope and stream, especially for humid regions such as our study catchment (Bracken and Croke 2007; Pavlin et al 2021). It is where both surface and subsurface flow pathways were connected continuously from hillslope to the outlet, leading to the nonlinearity in runoff generation (McGlynn and McDonnell 2003; McGuire and McDonnell 2010; Zhang et al 2021).

**4.2 Using XAI as an aggregation tool**

Similar classification based on soil moisture or its dynamics cannot provide comparable results with physical meanings at the watershed scale (Fig. S8). For classification based on SHAP values, if we look at the averaged soil moisture or its dynamics, the temporal variations were gradual and steady during the rising limb (Fig. 2). It would still be difficult to infer the occurrence of the abrupt rise in streamflow. It is only when we look at the average SHAP values of TR nodes that the abrupt rise could be observed to indicate the establishment of hydrologic connectivity and explain the nonlinearity in the hydrograph. This is because to upscale responses such as soil water movement, the pattern of spatial organization is needed, which is not accounted in simple averaging of point values (Blöschl and Sivapalan 1995; Viney and Sivapalan 2004). On the other hand, by calculating the importance of surrounding grids, SHAP values revealed the functional role of these grids, which can be considered as the combined results of both the point values and the underlying pattern of organization. This way, the average SHAP values have accounted for the pattern required for upscaling, making it comparable to the hydrologic response at watershed scales, i.e., the development of hydrologic connectivity.

The advantage of DL is that it is capable of learning from data, which may even beyond what we have known. Yet its 'black box' characteristic hinders it from facilitating scientific understanding. Using the hydrologic connectivity as an example, our work suggests that when applied at grid scale, XAI method could provide an alternative way to aggregate responses from point scale to watershed scale. By incorporating hydrologic knowledge, we can interpret the aggregated XAI results with physical meanings. Although the dataset used here is based on numerical simulations from a physically based hydrologic model, the development of hydrologic connectivity is not enabled in the model as a priori and is difficult to quantify from model outputs. The capability of XAI results in identifying both the timing and the abrupt rise in streamflow demonstrated the physical plausibility of this framework for cross-scale aggregation. Fine-tuning the model with in-situ measurements, further analysis could be conducted to scale up new findings to expand the applicability of our results as well as to advance our understanding of the physical reality.

**4.3 Limitations**

This study presented an illustrative example to demonstrate the capability of XAI methods to aggregate from small scale to watershed scale and gain scientific insights. The performance of the model is constrained by the quality and size of training data, which could also impact the accuracy in interpretation. This numerical experiment was conducted in a small watershed; therefore, the model's capacity of generalization is limited. Although this work is a theoretical study, applications to more diverse scenarios of rainfall and antecedent conditions could expand the range of the training dataset as well as the applicability of the results. The study area is a small humid mountainous watershed, to further explore the scaling capability, watersheds with a range of drainage area, climate and topographic characteristics will be needed. Moreover, in future work, incorporating physical constraints into the model structure directly can enhance the physical merits of the DL model. There is a small proportion of nodes on the hillslope that were classified as CH, which may be benefit from further refinement of the classification metrics and model prediction. Causal inference methods could also be used to investigate the causality between the XAI results.

**5 Conclusions**

Cross-scale aggregation of hydrologic responses is important for modeling and understanding, which requires both the aggregation of point values across an underlying organizational pattern. XAI results (i.e., SHAP values), by encapsulating the functional roles of the variables, can be considered as a combination of both the grid values and the underlying spatial organization, providing an alternative way for upscaling. Here we propose a new framework that integrates XAI with a priori hydrologic knowledge to aggregate point scale response to hydrologic behavior at watershed scale, using hydrologic connectivity, a watershed scale concept that is usually inferred indirectly from small scale measurements as an example. Our results demonstrates that the XAI results, i.e., the SHAP values, are able to differentiate the collective function of three sub-regions involved in runoff generation: the saturated area, the unsaturated area, and the riparian zone. The average SHAP values of the riparian area are able to reflect the development of hydrologic connectivity explicitly at the watershed scale, and can explain a nonlinear surge in streamflow. This proposed framework could be used to facilitate the upscaling of other responses for process understanding as well, for example, solute movement in the soil. The evaluation of functionality in model simulation could also help validate model plausibility and cross-model comparison (Müller et al., 2024). Causal inference methods would be used to help investigate the causality between the XAI results.

**Code availability**

The code and data used in the study are available at Li (2024).

**Author contribution**

YS: conceptualization, writing (review and editing), supervision, funding acquisition; LJY: methodology, formal analysis, data curation, writing (original draft preparation); Siva: writing (review and editing); RQH: conceptualization, writing (review and editing), supervision, project administration; LL and CYF: formal analysis. All authors worked on the manuscript.

**Competing interests**

The authors declare that they have no conflict of interest.

**Acknowledgements**

This research was supported by the Joint Funds of the Zhejiang Provincial Natural Science Foundation of China under Grant No: LZJWZ24E090002.

# Supporting Information for

# Explainable artificial intelligence (XAI) for scaling:

# An application for deducing hydrologic connectivity at watershed scale

Sheng Ye[1], Jiyu Li[1], Yifan Chai[1], Lin Liu[2], Murugesu Sivapalan[4,5], Qihua Ran[2,3*]

1Institute of Water Science and Engineering, College of Civil Engineering and Architecture, Zhejiang University, Hangzhou 310058, China

2Key Laboratory of Hydrologic-Cycle and Hydrodynamic-System of Ministry of Water Resources, Hohai University, Nanjing 210098, China

3National Cooperative Innovation Center for Water Safety & Hydro-Science, Joint International Research Laboratory of Global Change and Water Cycle, Hohai University, Nanjing 210098, China

4Department of Civil and Environmental Engineering, University of Illinois at Urbana-Champaign, Urbana, IL, USA

5Department of Geography and Geographic Information Science, University of Illinois at Urbana-Champaign, Urbana, IL, USA

**Contents of this file**

**Text S1. Control equation of soil water movement**

The InHM model define the movement of soil moisture in variably saturated soils based on Darcy's law and the Richards' equation. The InHM model employs a finite element method with control volume discretization. Variably-saturated subsurface flow in InHM is defined by Richards' equation as following:

$$\nabla \cdot f^a \vec{q} \pm q^b \pm q^e = f^v \frac{\partial n_s S_w}{\partial t}$$

where $\vec{q}$(m/s) is Darcy flux, $q^b(s^{-1})$ is a specified rate source/sink, $q^e(s^{-1})$ is rate of water exchange between porous medium and surface continua, $n_s$ is the porosity, $S_w$ is the saturation, $f^a$ is the area fraction associated with each continuum, $f^v$ is the volume fraction associated with each continuum, and $t(s)$ is time.

The Darcy flux is given by:

$$\vec{q} = -k_{rw} \frac{\rho_w g}{\mu_w} \vec{k} \nabla(\psi + z)$$

where $k_{rw}$ is the relative permeability, $g(m/s^2)$ is the gravitational acceleration, $\mu_w(kg/(m \cdot s))$ is the viscosity of water, $\vec{k}(m^2)$ is the intrinsic permeability vector, $\psi(m)$ is the pressure head, $z(m)$ is the elevation head. The predicted velocity is represented by S(t) - S(t-1).

The van Genuchten formula is applied to define the WRC and HCF, so water saturation ($S_w$) and relative permeability ($k_{rw}$) is defined by:

$$k_{rw}(S_w) = S_e^{1/2} \left[1 - \left(1 - S_e^{1/m}\right)^m\right]^2$$

$$S_w(\psi) = S_{wr} + (1 - S_{wr})[1 + |\alpha\psi|^n]^{-m}; \quad m = 1 - 1/n$$

where $\alpha$ and $n$ are the experience curve fitting parameters. Effective saturation $S_e$ is defined as:

$$S_e = max\left(0, \frac{S_w - S_{wr}}{1 - S_{wr}}\right)$$

where $S_{wr}$ is the residual saturation.

**Text S2. The explainable AI method**

The Expected Gradient (EG) method aims to calculate an importance score for each feature (e.g., precipitation or soil moisture) at each time step, where the absolute value of the score represents the magnitude of the feature's effect on the network output, commonly referred to as the level of importance. The EG method is derived from Integrated Gradients (IG), which calculate the integrated gradient that accumulates local gradients along a path from the selected baseline input x' to the target input x (Sundararajan et al, 2017). The path is usually simplified as a straight-line path (i.e., $\gamma(\alpha) = x' + \alpha(x - x')$) from the baseline input ($\alpha = 0$) to the target input ($\alpha = 1$). The IG for the $i$-th input feature is shown in the following equation:

$$\phi^{IG}(f, x, x') = (x - x') \times \int_{\alpha=0}^{1} \frac{\partial f(x' + \alpha(x - x'))}{\partial x_i} d\alpha$$

where $\partial f(x' + \alpha(x - x'))/\partial x_i$ denotes the local gradient of the network $f$ at a point interpolated between the baseline input and target input, $x$ is the input to be explained, which is part of the LSTM model's test set, and $x'$ serves as a background sample for reference, used to assess the impact of $x$ on the model's output.

However, using a single sample as the baseline leads to issues such as bias risk and insufficient generalizability. The EG method was developed to avoid specifying the baseline input (Erion et al, 2021). It assumes that the baseline input follows an underlying distribution $D$ derived from a background dataset (e.g., the training dataset), which is formulated as:

$$\phi^{EG}(f, x) = E_{x' \sim D, \alpha \sim U(0,1)} \left[ (x - x') \times \frac{\partial f(x' + \alpha(x - x'))}{\partial x_i} \right]$$

where $D$ typically denotes the uniform distribution defined over the entire training set, which is leveraged to help explain which features cause the output for $x$ to differ, on average, from the outputs at all the other samples in the training set. $U(0,1)$ denotes the uniform distribution over the interval from 0 to 1. In this study, $\phi^{EG}$ was calculated with the SHapley Additive exPlanations (SHAP) package (Lundberg & Lee, 2017) to quantify the importances of each feature in prediction.

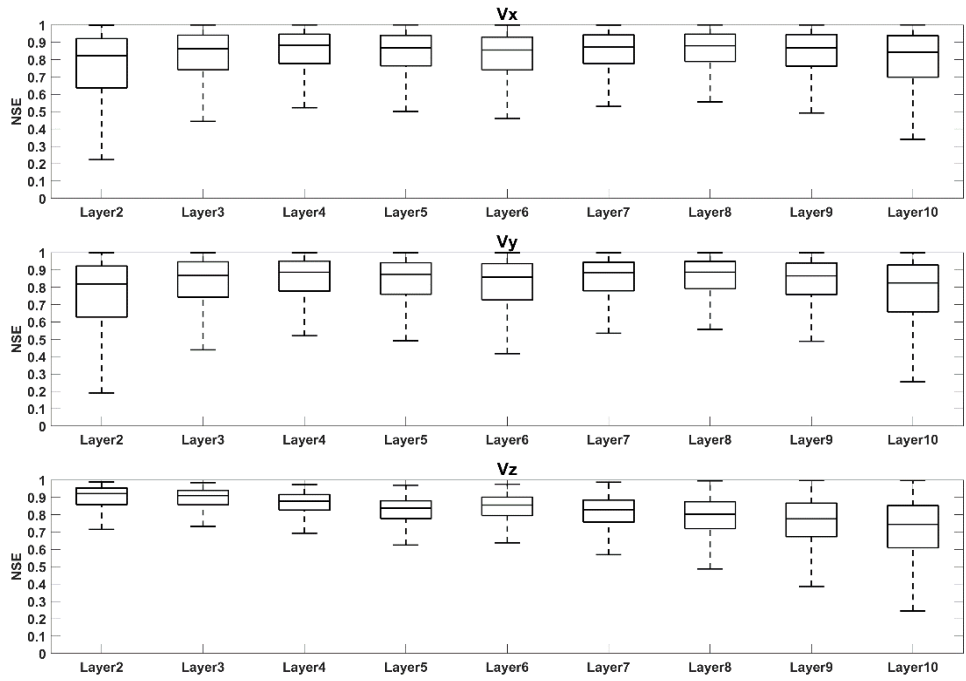

**Figure S1:** NSE calculated at different depths in the three directions, (a) Vx, (b) Vy, (c) Vz.

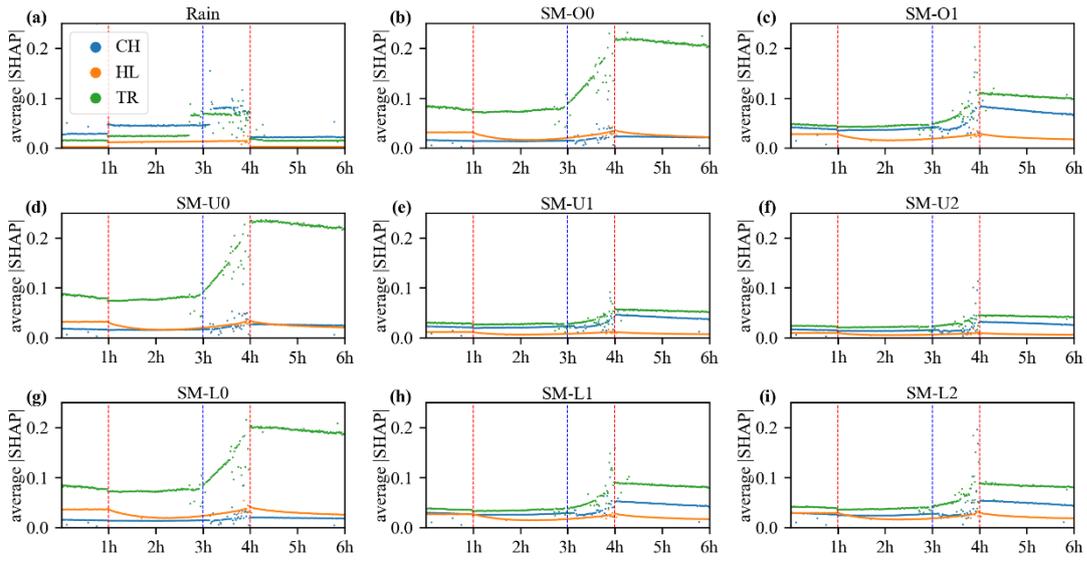

**Figure S2**: The average SHAP value sequences of rainfall and soil moisture at nodes with higher importance (O0, O1, U0, U1, U2, L0, L1, L2) in Y-direction of Layer 2.

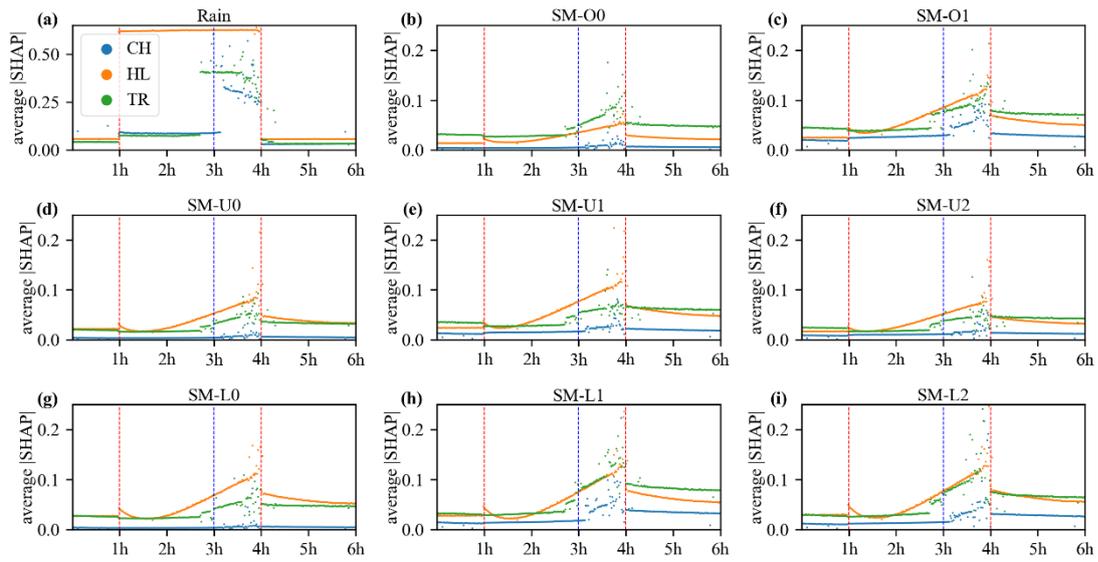

**Figure S3**: The average SHAP value sequences of rainfall and soil moisture at nodes with higher importance (O0, O1, U0, U1, U2, L0, L1, L2) in Z-direction of Layer 2.

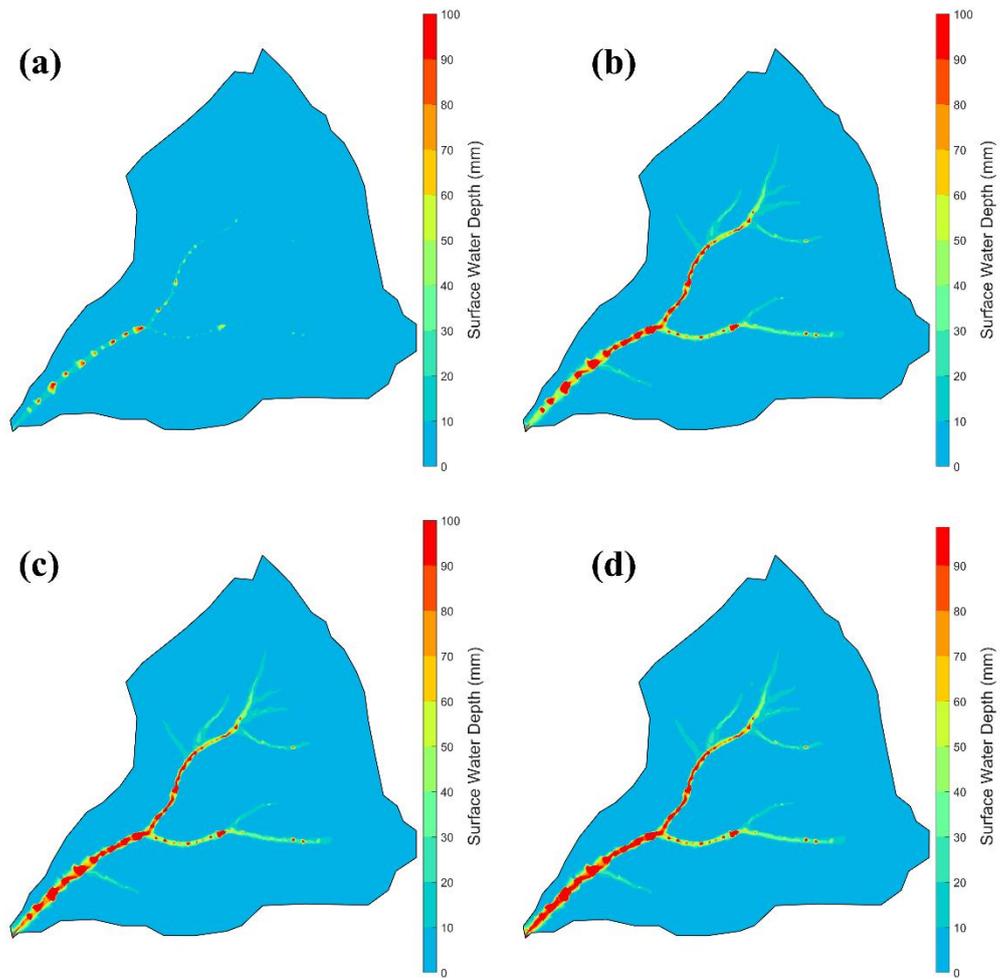

**Figure S4:** Spatial distribution of surface water depth at (a) start of rainfall; (b) before the abrupt rise; (c) after the abrupt rise; (d) end of rainfall.

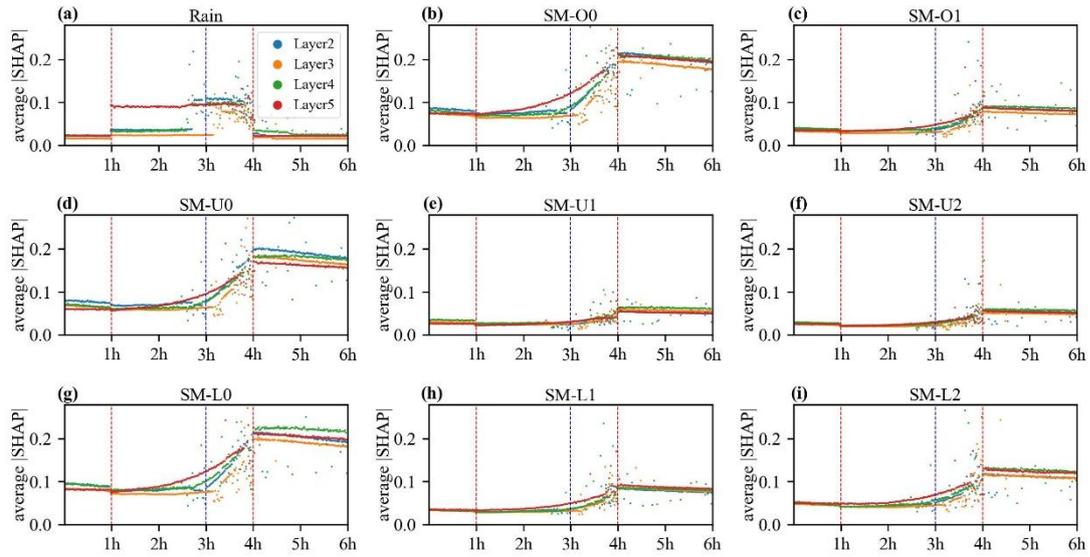

**Figure S5**: The average SHAP value sequences of rainfall and soil moisture at nodes with higher importance (O0, O1, U0, U1, U2, L0, L1, L2) in TR nodes at different soil depths in X-direction.

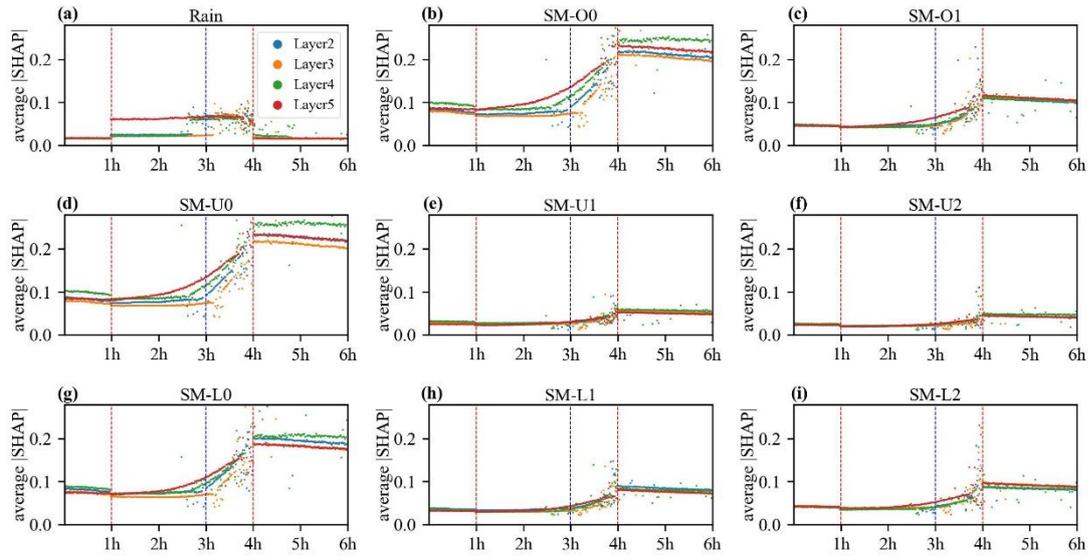

**Figure S6**: The average SHAP value sequences of rainfall and soil moisture at nodes with higher importance (O0, O1, U0, U1, U2, L0, L1, L2) in TR nodes at different soil depths in Y-direction.

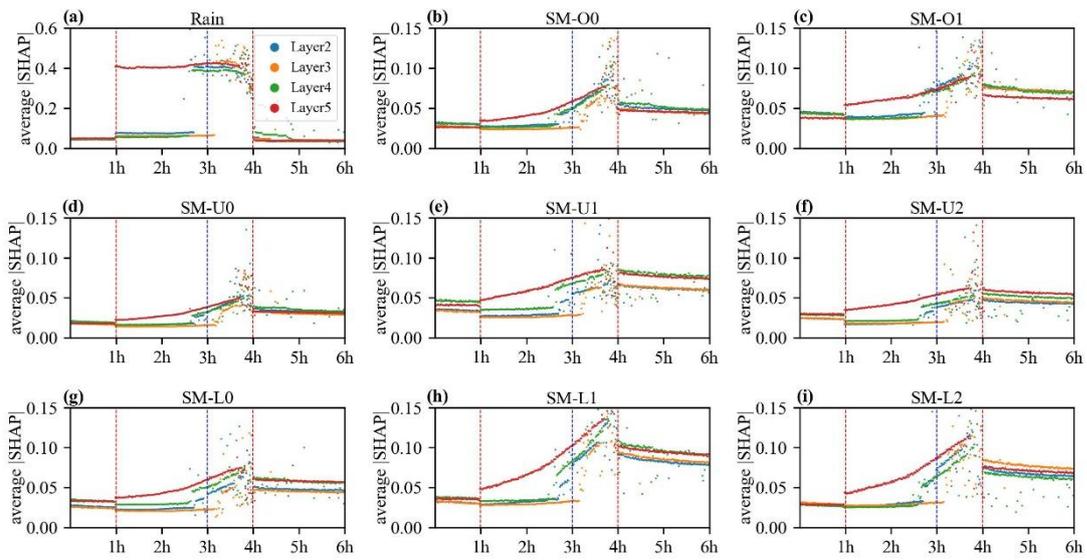

**Figure S7**: The average SHAP value sequences of rainfall and soil moisture at nodes with higher importance (O0, O1, U0, U1, U2, L0, L1, L2) in TR nodes at different soil depths in Z-direction.

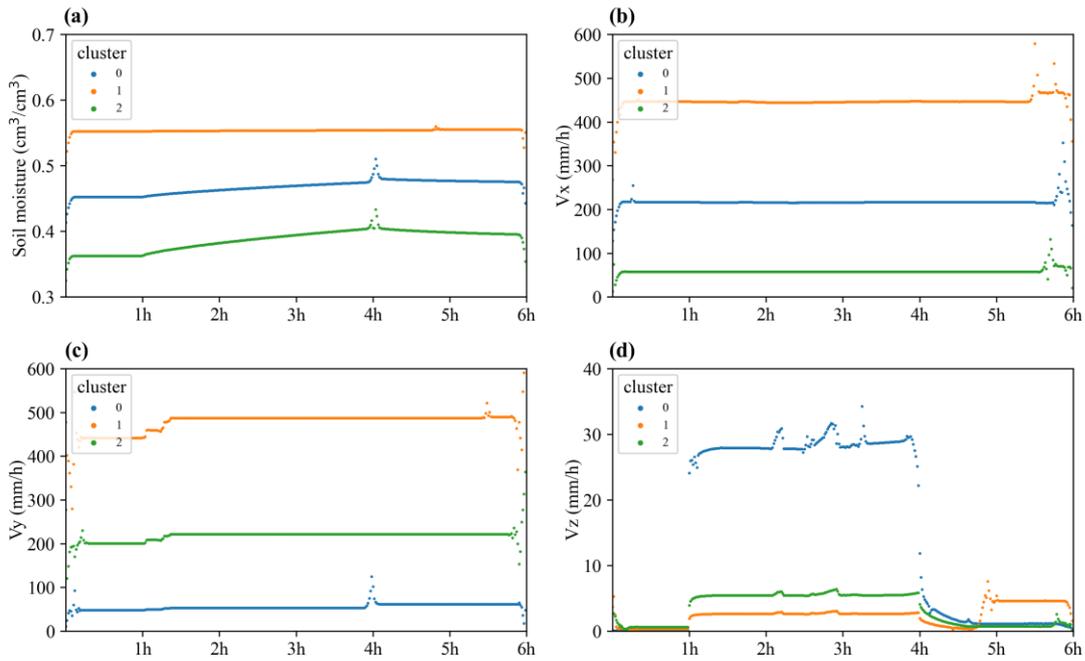

**Figure S8:** Clusters based on (a) soil moisture, (b) Vx, (c) Vy, and (d) Vz. Rainfall stars at 1h, ends at 4h, and the rapid rise in discharge occurs around 3h. As we can see, there is no abrupt rise in soil moisture, Vx, and Vy, while there is rise in Vz, it occurs at all three clusters for more than once, and drops quickly, different from the pattern in streamflow, and is difficult to relate it with rainfall-runoff processes.